\documentstyle[12pt]{article}

\newcommand{\be}{\begin{equation}}
\newcommand{\ee}{\end{equation}}
\newcommand{\bea}{\begin{eqnarray}}
\newcommand{\eea}{\end{eqnarray}}

\begin{document}

\baselineskip=14pt plus 0.2pt minus 0.2pt
\lineskip=14pt plus 0.2pt minus 0.2pt

\begin{flushright}
 hep-ph@xxx/9411261 \\
   LA-UR-94-3799 \\
\end{flushright}

\begin{center}
\Large{\bf
A CATCHING TRAP FOR ALL ANTIPROTON SEASONS} \\

\vspace{0.25in}

\large

\bigskip

Michael Martin Nieto\footnote{Email:  mmn@pion.lanl.gov}
and Michael H. Holzscheiter\footnote{Email:  mhh@lanl.gov}  \\
{\it
Theoretical$^1$ and Physics$^2$ Divisions\\
 Los Alamos National Laboratory\\
University of California\\
Los Alamos, New Mexico 87545, U.S.A. }

\normalsize

\vspace{0.3in}

{ABSTRACT}

\end{center}

\begin{quotation}

We describe the origin, development, and status of the Los Alamos
antiproton
catching trap.  Originally designed for the antiproton gravity experiment,
it
now is clear that this device can be a source of low-energy antiprotons
for
a wide range of physics, both on site, at CERN, and also off site.

\vspace{0.25in}

\end{quotation}

\vspace{0.3in}

\noindent {\it Ach ihr G\"otter$|$ gro{\ss}e G\"otter \\
In dem weiten Himmel droben$|$ \\
G\"abet Ihr uns auf der Erde \\
festen Sinn und guten Mut,}\dots \\
-- {\sc menschengef\"uhl}, Goethe

\newpage


\section{Introduction}

One of the many characteristics we have come to look forward to in
Herbert
Walther's work, is his  taking of a known technique and, with exciting
perception, his using it to produce wonderful new physics.
A very spectacular example was Walther's creation of ordered ion
structures  going around a ``race track" \cite{walt}; that is, a
radio-frequency, quadrupole, storage ring.

In honor of this spirit,
we wish to describe the development of
the Los Alamos, antiproton, catching trap, the uses of which may
may  go far beyond its original purpose.
As  reviewed in Section 2, the
catching trap grew out of the need for {\it some device} which would
slow down
antiprotons extracted from the Low Energy Antiproton Ring (LEAR), at
CERN.
These low-energy antiprotons would then be used
in an experiment to measure
the gravitational acceleration of the antiproton.
Ultimately, as detailed in Section 3, an elongated, cylindrical
Penning trap  was
devised, modified, and completed for this purpose.  As of
today, approximately
 $10^6$ antiprotons have been captured and cooled in it, after
having been extracted from one slow spill from LEAR.

However, having accomplished this, it is now clear to a wide community
that, serendipitously,  this device is a made-to-order source of new
particles:  low-energy antiprotons.  In  Section 4 we discuss
some of the many
uses to which this device may be put.  For example,
they can be used in  low-energy, antiproton,
nuclear and atomic physics experiments , as a storage vessel
for the creation of antihydrogen, and as
a source to fill small portable traps.  These traps  could then be sent
to universities and research institutions all over the world.


\section{The History of the Trap}

A decade ago,  no direct test of gravity had ever been performed on
antimatter.  This, and the lack of unambiguous evidence for the
``orthodox"
view that antimatter  experiences the same interaction as matter,
led to the suggestion that an experiment be performed at LEAR to
measure the
gravitational acceleration of the antiproton \cite{tgn}.

The idea was to cool antiprotons, in some manner, down to
approximately 4 K.  Then the antiprotons would be released  up  a
field-free ``drift-tube," similar in principle
 to the one Witteborn and Fairbank
used to measure gravity on electrons \cite{wf}.
After release, the hottest antiprotons would quickly go to the
top
of the drift tube, with the slower ones taking longer.  The cutoff time of
the
distribution would be when the least energetic of the antiprotons would
just
have enough energy to overcome the gravitational potential. This cutoff
time
is given by
\be
\tau = \sqrt{\frac{2L}{g}}~,
\ee
where $L$ is the field-free length of the drift tube.  As an example, for
normal gravity and a length of one meter, the cutoff time is
0.452 sec.
Statistically, about $10^6$ antiprotons must be sent up the drift tube to
obtain a value of $g$ for antiprotons relative to $g$ for negative
hydrogen ions to a few percent.

 This part of an actual experiment remains no small
feat, and continues to be  developed \cite{trap92,drift}.
The force of gravity is
equivalent to $10^{-7}$ V/m, which shows  to what extent field
variations,
like from the patch effect \cite{jordan}, must be overcome.  The
antiprotons
must also be launched in bunches, containing no more
 than approximately 100 particles, to
prevent  self-interactions from disturbing the measurement.  Thus,
the
idea is to store $10^6$ antiprotons in a small launching trap within the
drift
tube, and launch them 100 at a time by slowly dropping the voltage
trapping
the antiprotons.

But a first problem for the experiment was to decelerate the slowest
antiprotons from LEAR, with energies of approximately 5~MeV, to
4 K.
The original idea \cite{tgn} was to reverse the injection linac and use an
electrostatic
generator to slow down the particles.  By the time an official proposal
was
made \cite{beverini}, the choice was to use a radio-frequency
quadrupole
accelerator (RFQ) as a ``decelerator,"  after which  the particles entered a
Penning trap where they would be resistively cooled, and
eventually injected into the drift tube.

In the end, however, a number of factors led to the decision to
use    a very long Penning-style trap as the primary
 apparatus in which to store and cool the antiprotons from LEAR.
These factors included
the cost involved with an RFQ, the success of  experiment PS196
\cite{gabrielse} in trapping a smaller quantity of  antiprotons
in a small Penning trap by first using a simple foil  as a degrader, and
the results of direct foil
experiments  compared with TRIM model calculations \cite{trap92}.

The entire apparatus of the antiproton gravity experiment
is  schematically described in Figure 1.  With luck it will help answer
the exciting question of what is the
gravitational acceleration of antimatter \cite{ng}.
It may be interesting to note that now,  nearly a decade after the
original proposal was brought forward and after many arguments
have been made that gravity on antimatter should not be any
different from that on matter, it is becoming clear from the ``dark matter
problem" that we really do not understand
gravity even in most of  the normal universe, much less in the
antimatter world \cite{leap}.


\section{The Trap Today}

At CERN,
antiprotons are produced at very high energies.
Presently,   the lowest
energy at which they are delivered to physics users is
5.9 MeV, at the Low Energy Antiproton Ring (LEAR) \cite{h6}. To
further reduce this energy a number of methods have been proposed and
partially tested: Deceleration by an RFQ  operated in reversed mode
\cite{h7}, slowing down
antiprotons in a dilute gas in the ``anti-cyclotron"
\cite{h8}, and degrading the
antiproton energy by passing the beam through thin foils
\cite{h9,h10}.
The
degrading foil method is by far the simplest and cheapest of these
proposals.  During the past few years it has been used by two experiments at
LEAR with promising results
\cite{trap92,h2}. We therefore discuss this method in some
 detail.

When protons or antiprotons pass through matter they loose
energy by collisions with the nuclei of the material. If the thickness of
the material is increased, particles eventually are stopped in the material
and, in the case of antiprotons, annihilate. Using a Monte Carlo computer
code based on energy loss data for protons, one can calculate both the
expected linear density of material at which a maximum number of
particles with low energy is transmitted and also how large the number for a
specified energy bin should be. One finds this optimum thickness to be
near the point where 50\% of the incoming particles are transmitted.
These calculations predict that as much as 5\% of the incoming antiprotons
from LEAR
will be  transferred into an energy bin between 0 to 50 keV \cite{h10}.

At this energy
it is possible to electro-magnetically capture the antiprotons in a Penning
trap and to further reduce their temperature using electron cooling \cite{h15},
stochastic cooling \cite{h16}, or resistive cooling \cite{h17}.
Such a capture  has been successfully performed by experiment PS196 \cite{h2}.
They obtained a capture efficiency of $2 \times 10^{-4}$ per keV well depth.
More than
20,000 antiprotons were captured in a small Penning trap and
cooled to temperatures below 100 meV. Observed cooling times were
approximately 10 seconds. Energy widths as small as 9 meV were
directly observed by releasing the trapped antiprotons from the trap.

After the antiprotons are cooled, the trap well can be lowered again to
accept a new pulse of antiprotons into the same trap. This method of
``stacking" has been demonstrated by the PS196 team and approximately
100,000 cold antiprotons have been captured into their Penning trap,
utilizing about 10 consecutive pulses from LEAR.


\subsection{Trapping of 30 keV antiprotons from LEAR}

In order to determine the ultimate efficiency that can be achieved
for degrading and capturing antiprotons in a Penning-trap, the first part
of the PS200 experimental set-up was installed at LEAR
\cite{trap92}. This part
consists of a Penning type trap of 50 cm total length and 3.8 cm
diameter, situated in the horizontal, cryogenic bore of a
superconducting magnet capable of producing a magnetic field of up to 6
Tesla. Figure 2 shows the general layout of the ``test" experiment and the
different detectors used to monitor the incoming beam and to verify the
capture of antiprotons.

A particle pulse from LEAR is transported  to the front end of the
experiment.  After exiting the LEAR beam line through a 12
micron titanium window, the pulse passes through
a parallel-plate avalanche counter (PPAC)
for beam monitoring.  The beam then goes through a gas
cell, for fine tuning of the energy degrading, and enters the vacuum
system of the experiment through another 12 micron titanium window.
By then
the beam energy has been reduced to approximately 3.7 - 4.0 MeV. Due
to the transverse scattering
caused by the material the particles  pass
through, a relatively large angular spread is introduced into the beam.
But the beam can be focused by the fringe magnetic field of the
superconducting magnet.
In particular, by choosing a specific magnetic field strength for
the particular energy of the incoming beam, the focal point can be  placed
onto the entrance foil of the trap. In this 135 micron gold-coated
aluminum foil the antiprotons  loose more energy by collisions with the
atoms of the foil material.

Assuming proper adjustment of the
additional degrader material upstream, an optimum number of low
energy particles will exit from the downstream face of the foil. These
particles will be reflected by the electrical potential at the far end of the
trap and travel back towards the entrance electrode. This electrode is then
rapidly ramped up to potential before the particles can escape, thereby
capturing them within the volume of the trap.

The principal design parameter for the trap is the length
necessary to capture particles of energies up to 30 keV emerging from
the entrance
foil during a LEAR pulse of typically 200 ns duration. At a 1 m
round-trip distance, the time remaining after the last particle has entered
the trap before the first particle is reflected back to the entrance is 220 ns
for a 30 kV well depth. Our current 30 kV pulser has a 125 ns rise,
allowing a total of 95 ns for jitter and uncertainty in the trigger timing.

{}From these parameters we have constructed a trap structure
which consists of 7 electrodes: the entrance foil, a central region
comprised of five cylinders ( 2 endcaps, 2 compensation electrodes, and
the central ring), and a cylindrical, high-voltage, exit electrode.
The lengths and diameters have been carefully chosen to produce
a harmonic, orthogonalized, quadrupole potential in the central region
\cite{h18}.
For the purpose of the
initial antiproton capture, the trapping region is defined by the entrance
foil and the high-voltage exit electrode.
Except  for the small central  region,  the trap has no
harmonic properties and is the characteristic ``catching trap"
referred to
throughout this paper.

The central, harmonic region serves a dual
purpose: to initially hold cold electrons in preparation for the electron
cooling, and then to collect the cooled antiprotons after the electron
cooling has taken place. This
part of the trap is instrumented with two independent
tuned circuits to  detect  electrons and antiprotons via the
signals induced in the compensation rings.

To establish the capture of low-energy antiprotons, they are
released from the trap after a predetermined storage time.
The release is accomplished by
lowering the potential of the down-stream end-cap of the trap linearly
with  time, the time constant being
 large compared to the oscillation period of the
particles in the trap. Particles will escape from the trap when their
kinetic energy is greater than the potential barrier.

 The annihilation of antiprotons on the surface of the microchannel plate
detector (MCP) is detected by
using scintillators outside the magnet dewar as well as
by direct counts from the MCP. To reduce the background rate in the
``hot" accelerator environment, the scintillators are used in a 2-fold
coincidence set-up and can additionally be gated by the MCP pulses. The
detection efficiency has been deduced from Monte Carlo calculations
\cite{trap92}  to be approximately
7\%, a value which has been experimentally confirmed using
slow, continuous spills from LEAR.

This all generates a time-of-arrival spectrum which reflects the
energy distribution of the particles
in the trap prior to their release. Figure 3 shows such an energy spectrum
of approximately 500,000 antiprotons released from the trap 500 msec
after the pulser had fired to capture a pulse delivered from LEAR to our
experiment.

To measure the storage time of antiprotons in our trap, the delay
time between capture and release is varied. The number of detected
antiprotons for each of these  ``shots"
is normalized to the total intensity
reading from the NE110 beam monitor.  The results are plotted in
Figure 4.


\subsection{Cooling of antiprotons}

During earlier storage time measurements a noticeable change in the
spectral shape was noticed. After storage times of typically around
15 - 20 seconds, high-energy particles could no longer be observed and
the energy distribution had started to shift towards later channels in the
release spectrum without a decrease in the total number of particles.
After 30 - 40 seconds all counts in the arrival-time spectrum where
concentrated at energies below 1 keV. Figure 5 shows a selection of
energy spectra for 3 different storage times (8, 20, and 70 seconds).
Such a cooling time
would require an electron density of approximately $10^8$
electrons/cm$^3$
\cite{h15}.

As later tests revealed, electrons were
continuously produced by
field emission from sharp points on the trap electrodes. These electrons
were stored inside the well and cooled rapidly by synchrotron radiation.
In some recent experiments we installed an electron source, consisting
of a hot filament and appropriate extraction and focusing electrodes, in
the fringe magnetic field region.
Electrons produced by this source were trapped,
cooled by synchrotron radiation, and  collected in the central well.

Using both a resonant detection technique as well as extracting these
electrons from the central well
and counting them with the MCP, we established that we can
load the inner well with $10^8$ electrons using a primary electron beam of
50 mA for 10 - 30 seconds.
Antiprotons oscillating in the large catching trap interact via
Coulomb interaction with these electrons and dissipate energy into the
electron cloud.  It, in turn, is continuously cooled by synchrotron
radiation.  Finally,  both electron and antiproton clouds arrive at a thermal
distribution in equilibrium with the ambient temperature of the apparatus.

One of the main problems is the small overlap between the antiprotons
oscillating in the 50 cm long catching trap and the electron cloud
confined to the central region of the harmonic well. Standard electron
cooling calculations assume the two clouds to be completely
overlapping. A first-order approximation consists of diluting the
number of electrons, $10^8$, into the volume occupied by the antiprotons.
Under these conditions,
this effective electron density  is  only $2 \times  10^6$ e/cm$^3$.
{}From this density we calculate an initial time constant for
cooling of 140 seconds. This estimate is certainly only a lower limit,
since it does not account for the actual dynamics of the interaction
between the two clouds. The actual time constant could be higher
by a factor of ten. Accordingly, in a recent experiment, where we
carefully avoided any additional loading of electrons by corona
discharges, no cooling was observed.

Another source of the observed cooling could be collisions with the
residual gas. Assuming the main component of the residual gas to be
helium, the fractional energy loss of an antiproton per collision is 0.33.
Choosing a typical collision rate constant of
$2 \times 10^{-9}$ cm$^3$/sec,   we find
that an observed cooling time constant of approximately 20 seconds
would require a neutral density of
$8 \times 10^{7}$ cm$^{-3}$, or a pressure of $3 \times 10^{-11}$ Torr.
At the same time, such a residual gas pressure would result in
an annihilation-limited storage time of 100 to 1000 seconds, in
agreement with the observed storage times during our 1993 runs.

Since then
we have not only improved our control over the high voltage, but have
also installed an ``in-vacuum ultra-high vacuum valve" to separate the
cryogenic bore from the room temperature region of the vacuum
system. With this closed-off system we expect the residual gas pressure
to be significantly reduced.
According to the above estimates,
not having observed any significant cooling
at a delay time of 2000 seconds indicates that  the residual gas density
was less than $7.5 \times 10^{5}$ cm$^{-3}$,
corresponding to a pressure of less than $3 \times 10^{-13}$ Torr.

While both the above scenarios are possible, we do not have enough data to
distinguish between them. This will be part of the R\&D program for
upcoming runs. One way to attack this question consists of deliberately
spoiling the cryogenic vacuum by opening the ``in-vacuum" valve after a
set of data in the closed cryogenic configuration has been taken. In this
way we will be able to separate out the electron-density issues from the
residual-gas issues.


\subsection{ Ejection of antiprotons from the PS200 catching trap}

For most of the physics
experiments envisioned at this time, the antiprotons will
need to be ejected from the trap once the initial cooling has taken place.
For the gravity measurement proposed in PS200 all antiprotons will be
transferred in a single bunch into a small Penning trap at the bottom of
the vertical time-of-flight experiment. Here they will be resistively
cooled to 4.2 K and then released in bunches containing approximately
100 antiprotons each.

While a fast transfer of a single pulse is required for the gravity
experiment,  other experiments with ultra-low energy antiprotons
will require a `semi'-continuous beam, possibly with timing information
on the release of individual antiprotons. A number of possible schemes
can be conceived of to extract the cloud of antiprotons from the PS200
catching trap in this way.

Note that one can not just lower the potential at one end cap over an
extended period of time. Since all antiprotons will have been cooled to
an extremely low temperature, one would only obtain an extraction during
the very last fraction of the spill time. Instead, one can eject the
antiprotons by an evaporative process. Here the axial or cyclotron
resonance frequency of the stored antiprotons is weakly excited, leading
to a continuous heating and a slow ``boil-off' of particles from the well.
The rate of boil-off can be controlled by the amplitude of the
radio-frequency applied as well as by the detuning between the applied
frequency and the resonant frequency. Test experiments conducted at
Los Alamos using a smaller Penning trap filled with protons have
generated continuous spills of protons for approximately 30 minutes at a
time \cite{h19}.

The above  evaporative, slow spill can be used for experiments where a
low intensity of antiprotons and no timing information is needed.
But if a
time structure is required (e.g.,  for time-of-flight studies of the energy
loss in materials) a different method is proposed. This method was
originally developed to eject low-energy electrons from a Penning trap.
The time-of-flight of the ejected electrons
through an inhomogeneous magnetic field was then used to determine
 the electron magnetic moment
\cite{h20}.

The well depth was slowly reduced, allowing electrons to leave the
trap whenever their kinetic energy exceeds the well depth. Superimposed
on this linear ramp was a series of triangular spikes with a half width
longer than the oscillation period of the particles in the harmonic well.
During the time period of one of these pulses all electrons occupying the
energy band covered by the pulse amplitude were allowed to escape,
generating a micro bunch with a defined start time.

A derivative of this method was  used in the proton test
experiments at Los Alamos:
 A series of rectangular pulses, with a FWHM slightly
larger than the oscillation period of the trapped particles and an
amplitude of 1 - 2 Volts, was superimposed onto the constant trapping
voltage. Additionally, a weak RF drive was applied at the axial
resonance (or the cyclotron resonance) to continuously heat the particle
cloud. The amplitude of this drive was  such  that
continuous boil-off was not quite taking place. A multi-channel
analyzer with a 200 ns/channel time resolution was triggered with the
leading edge of each of these pulses and a time spectrum for 1000
individual pulses was obtained. The result (See Figure 6) was a pulsed
beam with a time width of 1.2 $\mu$sec and a repetition rate of
of 100 Hz  with one particle per pulse, on average.


\section{The Potential Uses of the Trap}

\subsection{Nuclear physics with ultra-low energy antiprotons}

The availability of low-energy antiprotons with a well-defined
energy has generated substantial interest amongst experimenters who
have studied low-energy antiproton phenomena over the past few years.
In this subsection we describe
 two specific examples and compare the  methods
currently used  to   alternative methods,
which are based on  our catching trap
and have  advantages.

The first group of experiments is a series of measurements on
energy loss and straggling of antiprotons passing through matter
\cite{h21,h22}.
These experiments were performed by passing the lowest-energy beam
available from LEAR (5.9 MeV) through a degrader material and then
using a time-of-flight tag to select the particles with a specific energy.

This method has distinct disadvantages at lower energies. If the thickness
of the degrader material is increased to reduce the beam energy below
approximately 1 MeV, both the energy spread and the angular spread
increase dramatically. The energy spread becomes equal to the mean
energy at approximately 1 - 2 MeV and the number of particles
available at a given energy decreases drastically below 1 MeV. Under
these conditions one can no longer speak of a ``beam of antiprotons." Not
only is the number of antiprotons available at the energy of interest
diminishing rapidly, requiring more and more integrated beam time from
the antiproton source to accumulate appropriate statistics, but also
the background due to ``unwanted" particles at higher energy
quickly becomes
overwhelming. These high-energy particles can annihilate in the
experimental set-up, producing false counts, and can even saturate the
detector system.

Here our catching trap could serve as a bunching system to
compress the phase space occupied by the antiprotons and to remove the
high-energy background from the measurements. By utilizing a 30 kV
well depth and by cooling the particles to less than 1 eV, one
could achieve an
enhancement of more than $10^4$ in energy density. This would allow
experiments to explore energy regimes far below the current limit of
10's of keV and to accumulate  much better statistics in the low-energy
region.

Using the PS200 catching trap, a well-defined
energy beam could be produced with a very small energy spread,
allowing the direct measurement of low-energy processes. A number of
such experiments have been proposed by the PS194 collaboration \cite{h24}.
Once approved, they could be performed over the
next few years with a much reduced impact on the LEAR operation.

A second group of experiments which would greatly benefit from
very low-energy, narrow-energy width, antiproton beams are those
requiring ultra-thin targets. One example is the study of the formation
and delayed decay of hypernuclei when antiprotons are stopped in thin
target foils. These processes were studied at LEAR by the PS177
collaboration
\cite{h25}. A shadowing method was used to distinguish between
prompt decays inside the target and delayed decays of hypernuclei
which had escaped the target. The lifetime of heavy hypernuclei in the
region of uranium was measured to be of the order of $10^{-10}$
 seconds. To
improve this method it would be desirable to use thinner targets, thus
allowing a larger fraction of the formed hypernuclei to escape. To
maintain a reasonable stopping rate in these ultra-thin targets a much
lower energy of the antiprotons would be required.

These again could be obtained from our catching trap. By using
the time structured extraction method described in the previous section
one could generate short micro bunches of antiprotons at energies from a
few hundred eV up to 30 kV.  One would  switch the
potential of the long, cylindrical, high voltage exit electrode (or another
electrode placed in the system for this specific purpose) during the time
the bunch is shielded from external potentials while inside this electrode.
The bunches would then be accelerated to the kinetic energy set by the
potential applied to this electrode upon exiting. Model calculations
indicate a near 100\% efficiency for this process and experimental tests to
characterize this beam structure are under way
\cite{h19}. As an additional
benefit the overall intensity of antiprotons entering the experimental
set-up would be  greatly reduced so the
background from annihilations outside the
target could be reduced almost to zero.

We currently are studying the possibility of measuring the structures of
neutron and proton ``halos" from prompt X-ray, Gamma-ray, and
annihilation particle emission. Low-energy antiprotons entering a nucleus
will preferably annihilate near the nuclear surface. The
resulting pions have a high probability of missing the rest of the nucleus,
thereby avoiding excessive excitation of the nucleus and subsequent
fission of the target. The result is a ``cold" daughter product with
proton and neutron numbers of either (Z-1, N) or (Z, N-1), depending
on whether the annihilation occurred on a neutron or proton. By
detecting a distinct signature for the two possible routes the neutron
distribution near the nuclear surface can be mapped out.

Jastrzebki  et
al. \cite{h26}  presented the
first measurements based on these ideas.  But
their experimental method is limited to  cases where both
daughters are radioactive, since off-line radio-chemistry methods are
used for the reaction analysis. Our experiment would not be limited to
radioactive annihilation products. Ultra-thin isotopically enriched targets
and prompt measurements would be used to take advantage of the low
energy properties of the extracted antiproton beam from the PS200
catching trap. Rather than studying the subsequent radioactive decay of
the daughter products we plan to observe the (prompt) de-excitation of
the daughters from the excited state to the ground state, {\it in situ}.

A first experiment will be using $^{48}$Ca.  This isotope is interesting for
two reasons.  Firstly, it decays into radioactive products in both
branches.  Therefore, it can be used to calibrate our prompt experiment
against the standard method.  Secondly, theoretical model calculations exist
\cite{hmad27}
which predict a significant difference in the neutron/proton
ratio at the nuclear surface between $^{48}$Ca and $^{40}$Ca,
making this a choice of significant theoretical interest.


\subsection{Antihydrogen production}

The simplest system which can be studied by atomic physicists
is the hydrogen atom  \cite{h28}. It is
the only one where theory can even attempt to find exact solutions, and
the one which has subsequently been studied with great precision and
success, both theoretically and experimentally.  Naturally, it is a
tantalizing dream that one might eventually be able to study
 its mirror image, the antihydrogen atom,
with the same precision.  Once could then
either
solidify or  expand our understanding of fundamental symmetries.
By stating this goal, we hence restrict ourselves to
possibilities that would yield antihydrogen in an experimental environment
suitable for precision measurements comparable to those achieved on the
hydrogen atom. Since antihydrogen will
always be an extremely rare object, one immediately realizes that it is a
sensible approach to cool and trap the antihydrogen atoms.

A variety of schemes for producing antihydrogen has been
proposed, and discussed in some detail
\cite{h29}-\cite{h43}. The first mentioning
of the possible production of antihydrogen in traps
was by Dehmelt and co-workers \cite{h43}.
For all practical purposes,
the schemes we describe below are those which
deserve close
attention by the trap community.


\subsubsection{Antihydrogen production using trapped plasmas}

This method was originally  proposed
 by  Gabrielse's group \cite{h33}. In a ``nested trap" scheme,
forming two Penning traps, the oppositely charged constituents
(antiprotons and positrons) for
antihydrogen production are held in separate clouds and cooled to 4 K,
or even lower \cite{h45}. At a definite time, the two clouds are merged by
lowering the electrostatic barrier between them, and antihydrogen is
formed. The rate constant for this process is strongly temperature
dependent and benefits vastly from cooling the particles.
While the rate can be extremely high (with $10^7$/cm$^3$ positron
density at 4.2 K one obtains
$\Gamma = 6 \times 10^6/$s), one specific problem must
be addressed. The antihydrogen atoms are  formed in highly
excited Rydberg states (n $\sim$ 100).  The atoms need to be quickly
de-excited,
before electrostatic field gradients from the Penning trap ionizes
them.
This de-excitation process needs to be carefully controlled
since it also effects the capability of traping the antihydrogen
atom once it is formed.


\subsubsection{Antihydrogen production by positronium-antiproton
collisions}

Alternatively, one can enhance
the radiative antihydrogen formation rate by several orders of magnitude
through coupling of the recombination process to a third particle.
This  increases the phase space constrained by
energy and momentum conservation. Such a proposal has been made
to create antihydrogen
utilizing collisions between positronium atoms and antiprotons
\cite{h38}. This
process can be interpreted as Auger capture of the positron to the
antiproton.  The cross sections have been estimated by Humberston,  et al.
\cite{h37}. They used charge conjugation and time reversal to link the cross
section
for positronium formation in collisions between positrons and hydrogen
to the antihydrogen formation cross sections.

Early calculations assumed
both $\bar{H}$
and Ps to be in the ground state and obtained a broad maximum
in the cross section of $3.2 \times 10^{-16}$ cm$^2$ at
an antiproton kinetic energy of
approximately 2.5 KeV. Calculations of the total $\bar{H}$-formation
cross section using classical and semi classical methods \cite{h47}
have obtained values of
$\sigma(\bar{H})$ which are considerably larger than the
ground-state results. Values for the formation of $\bar{H}$ in excited states
are
given by Ermolaev,
 et al. \cite{h48} and indicate that there is a large cross
section to low-lying excited $\bar{H}^*$ states, which therefore would be
directly  accessible to spectroscopic studies.

Charlton \cite{h40} has discussed the formation of excited $\bar{H}^*$
atoms via collisions between antiprotons and excited positronium. The
cross section follows a classical
$(n_{Ps})^4$ scaling, where  $n_{Ps}$ denotes the
principal quantum number of the positronium atom, leading  to large
enhancements in the reaction rate. This process can also be utilized to
preferentially populate specific low-level excited states for spectroscopic
purposes.


\subsection{Precision measurements using antihydrogen}

Considering the effort necessary to produce antihydrogen one
must  ask  what further physics benefits  such an
endeavor could yield. In principle, these can be found in two areas.
A comparison of the results of spectroscopic measurements of hydrogen and
antihydrogen  would constitute a test of CPT at a level rivaling
even the result on the kaon system.  The study of the gravitational
interaction of antimatter with the Earth's gravitational field
would test the
validity of the weak equivalence principle (WEP) and possibly shed
light on the problem of unifying gravity with the three other forces.

The precision of spectroscopic studies of hydrogen advanced
enormously over the last decade. Today the highest precision has been
achieved for the hyperfine structure ($6.4\times 10^{-13}$) and for the
1s-2s level difference ($1.8\times 10^{-11}$), from which one
obtains the Rydberg.
Based on the lifetime
of the 2s state of 1/8
second and the natural linewidth connected to this, a precision of
$10^{-18}$ for the measurement of the Rydberg
has been speculated as being possible.
This latter precision will  most likely require
using trapped hydrogen atoms, an environment which would be directly
applicable to antihydrogen.


\subsubsection{CPT invaniance}

Currently the best tests of CPT invariance
have been performed in the kaon system followed by precision
comparisons of the magnetic moments and masses of electron, positron,
proton, and antiproton. The comparison of the inertial masses of
the proton
and the antiproton has now reached a precision of
$1.4 \times 10^{-9}$   \cite{h63}. In the
strict sense this must be considered only a measurement of the ratio of
the charge-to-mass ratios of the two particles. It has been proposed
    \cite{h64} that by
combining the direct determination of the cyclotron frequencies with the
measurement of the Rydberg of protonium
one could extract an independent CPT
test.  But with  the current precision on the Rydberg
\cite{h65}, a CPT test of  only $2 \times 10^{-5}$ is possible. Using the
Rydberg of antihydrogen, one could
construct a limit for the charge equality between antiproton and proton
which is entirely based on frequency measurements, and could therefore
yield a direct test of CPT at a level of $10^{-11}$ .


\subsubsection{Gravity on antimatter}

Often the arguments are made that measuring gravity on charged
antimatter is nearly impossible due to the interaction of the charge with
stray electric fields and that it would be advantageous to use a neutral
particle instead. To compare  charged and
uncharged experiments  in a fair way one needs not only
to consider the added complication of producing the antihydrogen
atoms but also one must devise a possible method of measuring the
gravitational acceleration with sufficient precision.

Although they  are repeatedly  discussed, purely ballistic
methods to measure gravitational acceleration on antihydrogen atoms
can be ruled out. Even if  antihydrogen atoms could be laser cooled to
the photon recoil limit of
$T = \Gamma\hbar/2k = 2.4$ mK, this temperature
would still correspond to a distribution of height of approximately 1 m in
the gravitational field. A precise determination of the centroid of a cloud
of this dimension is not possible. Similarly, if one should be able to
generate an atomic fountain with a mean energy of the photon recoil and
a spread of  half that value, the observed time-of-flight over a height of
10 cm will be ($14 \pm 7$) msec, again not yielding a precision
measurement of g. The only hope in the latter case would be to perform
an end point measurement similar to the PS200 proposal, but this time
with more particles near the end point.

A potentially much more powerful method could be developed
based on the work of  Chu and collaborators \cite{h66}. In their
experiment they  used velocity-sensitive, stimulated Raman
transitions to measure the gravitational acceleration, $g$, of laser-cooled
sodium atoms in an atomic-fountain geometry.

In their method an ultra-cold beam from
an atomic trap is launched upwards and is subjected to three subsequent
pulses to drive a two-photon Raman transition between the
 F = 1 and F = 2 sublevels of
 the 3S$_{1/2}$ state. A Raman transition is used to provide a large
photon recoil velocity while
still satisfying the metastability of the states
necessary for long interaction times. A first ($\pi/2$) pulse prepares the
sample in a superposition of the  $|1,p\rangle$ and the
$|2,p + \hbar k\rangle$ states. The
second ($\pi$) pulse  reverses the populations and a third
($\pi/2$) pulse
causes the wave packets to interfere. The interference can be
detected by probing the number of atoms in state $|2\rangle$.
In the absence of
any external forces acting on the atoms, the final state of an atom will
depend on the phase of the driving Raman field.

This result can be
extended to an atom falling freely in the gravitational field. In the frame
of reference falling with the atom, the Raman light fields appear Doppler
shifted linearly in time, which shows up as a phase shift varying as the
square of the time:
\be
\Delta \phi = - ( k_1 - k_2) gT^2 ,
\ee
where k is the wave number of the Raman light field and g
is the gravitational acceleration.  Since the Doppler shifts
($\sim 2k_1gT$) are much larger than the Rabi frequency, an active frequency
shift between the three subsequent pulses
must be used to compensate for the
deceleration of the atoms in the fountain.

Using a 50 ms delay between the pulses, distinct interference
fringes were observed and a least square fit to the data gave an
uncertainty in the phase determination of $3 \times 10^{-3}$ cycles.
This  represented a sensitivity to g of
$\Delta g/g  = 3 \times 10^{-8}$. A higher sensitivity is
expected to be obtainable when cesium is used instead of sodium because
of a large reduction of the rms velocity spread. This current work was
done with 30 K sodium atoms (representing a 30 cm/s velocity spread).
For  cesium  one may expect an rms spread of only 2 cm/s.
Therefore,  a much larger portion of the sample will be contributing to the
fringes.

A translation of this method to the hydrogen/antihydrogen
case will not be trivial or straight forward.
Hydrogen is (and antihydrogen certainly would be even more) ill
suited for high precision measurements, notwithstanding the enormous
    advances in hydrogen spectroscopy over the last years \cite{h67}.  A large
problem will be imposed by the
much higher photon recoil limit for laser cooling hydrogen atoms ($\sim$ 3
mK), which gives an rms velocity spread of approximately 700 cm/sec. A
much faster fountain beam, resulting in greatly increased experimental
dimensions, will have to be used. Therefore, a much larger fraction of
the initial beam pulse will be lost due to ballistic spreading during the
flight time of the sample and much less than 1\% of the initial population
can be expected to contribute to the fringes. This will also cause severe
problems for the antihydrogen/hydrogen comparison, since the supply of
antihydrogen atoms will be limited to small numbers and a re-trapping
scheme needs to be incorporated into the experiment. Nevertheless, this
method is the only one identifiable in the current literature which shows
the potential of a high-precision measurement of g on antihydrogen
atoms.


\subsection{Portable traps}

A final application of the catching trap would be to develop  smaller,
portable traps into which antiprotons could be ``decanted."
Such portable traps are actually prototyped by the launching trap of the
antiproton gravity experiment.  A current prototype is of approximate
length 8 cm, and protons have been  resistively cooled and stored
in it for hours.  A totally self-contained unit, with  magnet
and  cryogenics, could be of height 1 m,  diameter around 30 cm, with a a total
operational weight of approximately 100 kg.

With such a trap, one could envision a number a things.  Firstly, every
university could ``bring the mountain to Mohammed."   For both small-scale
research and education, each university could have its own supply of
antiprotons.  When the supply was gone, a new shipment could be obtained.
There are also possible medical applications, such as
producing short-lived medical isotopes, such as O$^{15}$, on site.
Lastly, of course, it may be a better idea to bring the antiprotons for
antihydrogen studies to outside laboratories, with their lasers and
magnetic traps, rather than to try to bring this equipment to the hostile
environment of an accelerator floor.

Ironically, in the end
the biggest problem may be obtaining import licenses for antimatter.


\section{Discussion}

	Recent advances in trapping and cooling of antiprotons have
opened up new opportunities for research with ultra-low energy
antiprotons.  Already now significant improvements in testing CPT in
the baryon sector have been accomplished, but the ultimate precision
will only be possible once antihydrogen can be formed and contained.
Long storage times of antiprotons and protons as well as recent advances
in trapping and cooling of neutral hydrogen bring this once futuristic
idea into the realm of the technically possible, even though experimentally
challenging. A number of crucial steps still need to be taken. But in the
meantime the possibility of a wide spectrum of interesting nuclear,
atomic, and gravitational physics can be pursued.

Finally, having come full circle, we hope that the exciting physics
that is now possible, with captured and cooled antiprotons, reflects
well on the spirit that Herbert Walther has shown in his work.
{\it Most of all, we wish to join in on the admiring congratulations to
Professor Walther on his 60th birthday.}



\newpage

\section*{Figure Captions}

Figure 1.  A schematic  diagram of the gravity experiment. \\

\noindent
Figure 2.  A diagram of the ``catching trap" and the associated
cryogenics, magnet, dewar,  degraders, and detectors. \\

\noindent
Figure 3.   An energy spectrum
of approximately 500,000 antiprotons released from the trap. \\

\noindent
Figure 4.   The number of antiprotons remaining in the
trap as a function of time.  \\

\noindent
Figure 5.  A selection of
energy spectra for 3 different storage times (8, 20, and 70 seconds).
This demonstrates cooling. \\

\noindent
Figure 6.  Time structure of a pulsed proton beam extracted from
our Penning trap.  The horizontal axis gives the time.

\end{document}